# Spin-orbit torques in bulk collinear antiferromagnets: complete classifications and the induced spin dynamics

Yizhuo Song[1], Jianting Dong[1], Jiahao Shentu[1] and Jia Zhang[1*]

[1]*School of Physics and Wuhan National High Magnetic Field Center,*

*Huazhong University of Science and Technology, 430074 Wuhan, China*

*jiazhang@hust.edu.cn

## Abstract

Electric field induced spin-orbit torques (SOTs) are the crucial mechanism for electric regulations of antiferromagnetic order. However, the spin-orbit torques in antiferromagnets (AFMs) and the induced spin dynamics remain largely unexplored. In this work, the full classifications of SOTs in bulk collinear AFMs have been achieved based on magnetic point group. Dependent on the symmetries connecting the opposite spin sublattices, the SOTs are classified into six distinct types. Among them, the SOTs and the induced Néel vector dynamics in three representative AFMs have been investigated, where the spin sublattices are connected by fractional translation, spatial inversion, and neither by translation nor inversion symmetry respectively. The SOTs on spin sublattices have been calculated by first-principles calculations based on Kubo linear response theory, and then the induced spin dynamics are simulated by Landau-Lifshitz-Gilbert equations. In typical *PT* symmetric AFM and the inversion symmetry breaking altermagnet, the simulations indicate that the deterministic switching of Néel vectors can be driven by field like torques. What's more, the fully electric writing of multiple antiferromagnetic domains into single domain state with preset Néel vector direction and 180° deterministic switching may also be realized. Our work may shed light on the current control of antiferromagnetic orders in collinear AFMs. Especially, for inversion symmetry breaking altermagnet, the electric writing and reading of Néel vector are highly desirable for antiferromagnetic memory applications.

# I. Introduction

Current induced spin-orbit torque (SOT) leads to the reorientation of the magnetic order. Considerable research has previously been conducted on the generation of spin torque and the process of its dynamics in ferromagnetic systems, leading to the development of magnetic random-access memory (MRAM)[1]. However, ferromagnets are inferior to antiferromagnets in terms of the reading and writing speed of information, storage density, energy consumption and immunity to external magnetic field interference. Meanwhile, the possibility of combining electrical manipulation and detection of antiferromagnetic order have been experimentally demonstrated. This indicates that antiferromagnets may be promising for information storage in electronic memory devices, opening new pathways for fundamental research on the spin torque in antiferromagnets and the induced spin dynamics.

The effective control of antiferromagnetic orders is therefore the primary focus of the antiferromagnetic spintronics. Železný et al. first predicted that electric switching of magnetic moments induced by staggered Néel fields in *PT* symmetric bulk antiferromagnets[2][3], which was demonstrated by the experiments[4][5]. Bulk AFMs can inherently possess spin-orbit torque, enabling the switching of Néel order and potentially allowing for reading and writing of information independently. Subsequently, the analysis of the symmetry of the spin-orbit torque in locally and globally noncentrosymmetric crystals[6], has proved the damping like and field like torque do exist respectively, which lead to efficient manipulation on antiferromagnetic moments. Based on the theoretical predictions, the in-plane switching of the Néel vector in the antiferromagnetic $Mn_2Au$ by current pulse density of $10^7$ $A/cm^2$[7][8] and terahertz writing speed in the antiferromagnetic CuMnAs[9] were realized by using intrinsic spin-orbit torques.

However, the systematic analysis and full classifications of SOTs in bulk AFMs have remained limited. Especially, the newly discovered spin-splitting AFMs with spin-polarization in momentum space enrich the material candidates with fascinating spin transport properties[10]. Besides, in contrast to bulk noncollinear AFMs [11][12], bulk collinear AFMs do not have non-relativistic spin torque[13]. The mechanism of SOT and the induced spin dynamics in bulk collinear AFMs remain ambiguous.

In this work, the spin-orbit torque in collinear AFMs in the absence of external spin current source has been classified into various types based on magnetic point groups (MPGs). The spin torque on spin sublattices in collinear AFMs have been calculated by using first-principles calculations in representative AFMs. Additionally, the spin dynamics of Néel vectors in AFM candidates under current induced flied like and damping like torque will be investigated.

## II. Symmetry classifications of electric field induced magnetic field and spin-orbit torque in collinear AFMs.

The spin torque can only exist in the presence of spin-orbit coupling (SOC) and there is no non-relativistic spin torque in collinear AFMs. Without SOC the spin is a good quantum number and the electric induced non-equilibrium spin density aligns with local magnetization and result in zero torque. The non-relativistic spin torque may only exist in some non-collinear AFMs.

To investigate the spin torque in AFMs, one mush introduce electric field induced magnetic field and spin torques acting on each spin sublattice. Based on the linear response theory, the induced magnetic field $B_E$ and torque $T_E$ on each spin sublattice can be written as[6]:

$$B_E = \chi E \; ; \; T_E = t E \quad (1)$$

Where $E=(E_x, E_y, E_z)$ is the electric field, $\chi$ is linear response 3×3 matrix for magnetic field, and $t$ is 3×3 torkance matrix. The induced torques and fields are correlative and connected by $T_E = M \times B_E$, where $M$ is the magnetization.

The electric field induced magnetic field and torque can be labeled as time reversal odd ($T$-odd) and even terms ($T$-even) on spin sublattices and transform differently under symmetry operations in magnetic point group. For instance, the magnetic field $\chi$ matrix should be transformed as follows[6]:

$$\chi_{A'}^{\text{even}} = \det(R) R \chi_A^{\text{even}} R^{-1} \; ; \; \chi_{A'}^{\text{odd}} = \pm \det(R) R \chi_A^{\text{odd}} R^{-1} \quad (2)$$

Where *R* is the magnetic point group (MPG) symmetry operation of a given AFMs which connects magnetic site A and A' (A' could be A or B lattice with the same or opposite magnetic moment), and ± in equation (2) is for *R* does not contain and contains time reversal operation *T* respectively.

Among 122 MPGs, except 16 cubic MPGs there are in total 106 MPGs that can describe collinear AFMs[14]. The spatial inversion *P* is the principle symmetry operation that determines the induced magnetic field and torque on spin sublattices of AFMs. Based on different combinations of *P*, *T* and *PT* symmetries, the 106 MPGs for describing collinear AFMs can be further categorized into six types as listed in Table I.

For type I AFMs with *T* but without *P* symmetry in MPG, the opposite spin sublattices A and B can be connected by fractional translation *τ*. By following the transformation in eq.(2), the induced *T*-even and *T*-odd magnetic field tensor *χ* and torkance *t* for two opposite spin sublattices A and B are connected by the following relations:

$$\chi_B^{even} = -\chi_A^{even}; \chi_B^{odd} = \chi_A^{odd} \quad (3)$$

$$t_B^{even} = t_A^{even}; t_B^{odd} = -t_A^{odd}$$

There will be opposite onsite odd torque and same even torque on A and B, which leads to net *T*-even torque.

For type II AFMs (Mn$_2$Au, CuMnAs) without *T* and *P* symmetry, but have *PT* symmetry in MPG, the opposite spin sublattices *A* and *B* are connected by spatial inversion. The tensor *χ* and *t* for two spin sublattices are connected by:

$$\chi_B^{even} = \chi_A^{even}; \quad \chi_B^{odd} = -\chi_A^{odd} \quad (4)$$

$$t_B^{even} = -t_A^{even}; t_B^{odd} = t_A^{odd}$$

There will be opposite onsite *T*-even torque and the same *T*-odd torque, which leads to net *T*-odd torque.

Type III AFMs with breaking *T* and *PT* symmetry in MPG (without *Tτ* and *PTτ* symmetries in MSG) are spin-splitting AFMs[10]. Type III AFMs can be further classified into three different types, *i.e.* type III-1, type III-2, and type III-3. Type III-1 AFMs are centrosymmetric spin-splitting AFMs with magnetic ions as inversion center.

There will be zero onsite and net torque. Since if magnetic ions are an inversion center, the inversion symmetry $P$ connects two identical spin sublattices and lead to the following relations:

$$\chi_{A,B}^{even} = -\chi_{A,B}^{even}; \chi_{A,B}^{odd} = -\chi_{A,B}^{odd} \quad (5)$$

There will be vanishing onsite spin-orbit torque on each sublattices.

Type III-2 AFMs are centrosymmetric spin-splitting AFMs but the magnetic ions are not inversion centers. The field induced magnetic field $\chi$ for two spin sublattices A and B that are inversion pairs described by the following relations:

$$\chi_A^{even} = -\chi_B^{even}; \chi_A^{odd} = -\chi_B^{odd} \quad (6)$$

$$t_A^{even} = -t_B^{even}; t_A^{odd} = -t_B^{odd}$$

Therefore, type III-2 AFMs have opposite onsite $T$-even and $T$-odd torques, which leads to zero net torque.

Type III-3 AFMs are spin-splitting AFMs without inversion symmetry $P$, the opposite spin sublattices A and B cannot be connected either by fractional translation $\tau$ or spatial inversion should have onsite as well as net $T$-odd and $T$-even spin torques.

For Type IV AFMs with both inversion $P$ and $T$ symmetry operations in MPGs, the opposite spin sublattices A and B can be connected by fractional translation $\tau$ as well as spatial inversion $P$. By satisfying eq.(3)-(4) simultaneously, there will be zero onsite and net spin torque.

TABLE I. Classification of 106 MPGs for describing collinear AFMs based on $T$, $P$ and $PT$ magnetic point group symmetries. The number in parenthesis in the first column is the numbers of MPGs belonging to the types of collinear AFMs. Please note the band structures for type II and type IV with $PT$ symmetry are spin degenerate. Type I AFMs may have SOC induced spin-polarization, while type III AFMs exhibit non-relativistic spin splitting band structures.

| Type | MPG | $T$ | $P$ | Connections of opposite spin sublattices | $PT$ | SOT (net) | SOT (onsite) | Candidates |
|---|---|---|---|---|---|---|---|---|
| I (18) | $11', 21', m1', 2221',$ $mm21', 41', \bar{4}1', 4mm1'$ $4221', \bar{4}2m1', 31',$ | ✓ | × | translation ✓, inversion × | × | nonzero | nonzero | MnS$_2$, MnPt bilayer |
| | $321', 3m1', 61', \bar{6}1',$ $6221', 6mm1', \bar{6}m21'$ | | | | | | | |
| II (18) | $\bar{1}', 2'/m, 2/m',$ $m'mm, m'm'm', 4/m',$ $4'/m', 4/m'mm,$ $4'/m'm'm, 4/m'm'm',$ $\bar{3}', \bar{3}'m, \bar{3}'m', 6'/m,$ $6/m', 6/m'mm,$ $6'/mmm', 6/m'm'm'$ | × | × | translation ×, inversion ✓ | ✓ | nonzero | nonzero | Mn$_2$Au, CuMnAs |
| III-1 | $\bar{1}, 2/m, 2'/m', mmm,$ $4/m, 4'/m, 4/mmm,$ $4'/mm'm, 4/mm'm', \bar{3},$ | × | ✓ | magnetic ions are inversion center | × | zero | zero | CrSb, RuO$_2$, KV$_2$Se$_2$O |
| III-2 (17) | $\bar{3}m, \bar{3}m', 6/m, 6'/m',$ $6/mmm, 6'/m'mm',$ $6/mm'm'$ | × | ✓ | magnetic ions are not inversion center | × | zero | nonzero | $\alpha$-Fe$_2$O$_3$ |
| III-3 (44) | $1, 2, 2', m, m', 222,$ $2'2'2, mm2, m'm2',$ $m'm'2, m'm'm, 4, 4',$ $\bar{4}, \bar{4}', 422, 4'22', 42'2',$ $4mm, 4'm'm, 4m'm',$ $\bar{4}2m, \bar{4}'2'm, \bar{4}'2m',$ $\bar{4}2'm', 3, 32, 32', 3m,$ $3m', 6, 6', \bar{6}, \bar{6}', 622,$ $6'22', 62'2', 6mm,$ $6'mm', 6m'm', \bar{6}m2,$ $\bar{6}'m'2, \bar{6}'m2', \bar{6}m'2'$ | × | × | translation ×, inversion × | × | nonzero | nonzero | KV$_2$SeTeO |
| IV (9) | $\bar{1}1', 2/m1', mmm1',$ $4/m1', 4/mmm1', \bar{3}1',$ $\bar{3}m1', 6/m1', 6/mmm1'$ | ✓ | ✓ | translation ✓, inversion ✓ | ✓ | zero | zero | L1$_0$-MnPt |

The corresponding AFM candidates belonging to different types are also listed in Table I. Fig.1 shows the crystal and antiferromagnetic structures for several type III-1 AFMs (centrosymmetric altermagnets with magnetic ions as inversion centers include CrSb, RuO$_2$, KV$_2$Se$_2$O etc.) with zero onsite SOTs and type IV AFM L10-MnPt. For type IV AFMs, represented by L1$_0$-MnPt, possess *P*, *T*, and *PT* symmetries. This symmetry combination constrains both the onsite and net torque must be zero.

In the following, we will focus on three representative AFMs including type I MnPt bilayer, type II Mn$_2$Au, and type III-3 KV$_2$SeTeO, while the type III-2 AFMs with limited metallic material candidates and no favorable current induced spin dynamics could occur in the presence of SOT (as we will discuss in the dynamics section of MnPt bilayer) will not be discussed in details. The SOT in those AFMs will be evaluated by first-principles calculations and the Néel vector dynamics will be investigated.

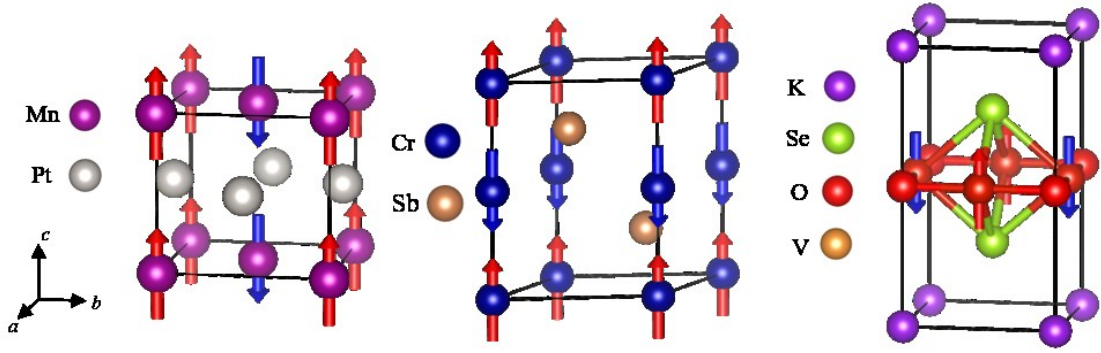

Fig. 1. The crystal structures of AFMs without onsite SOT for $L1_0$-MnPt, CrSb and $KV_2Se_2O$. The MSG of $L1_0$-MnPt with lattice constant of $a=7.521\ a_0$, $c=7.011\ a_0$ is P4/*mmm*. The MSG of CrSb with lattice constant of $a=7.799 a_0$, $c=10.301\ a_0$ is P6$_3$/*mmc*. The MSG of $KV_2Se_2O$ with lattice constant of $a=7.457\ a_0$, $c=13.818\ a_0$ is P4/*mmm* ($a_0$ is the Bohr radius). The magnetic moments aligned along the [001] direction.

## III. The torkance for representative AFM materials

As shown in Fig.2, the magnetic easy axes for MnPt bilayer, $Mn_2Au$, and $KV_2SeTeO$ are determined to be along [100], [110], [001] and the corresponding equivalent directions, respectively (please refer to the subsequent section for magnetic crystalline anisotropy calculations). The torkances in MnPt bilayer, $Mn_2Au$ and $KV_2SeTeO$ by setting magnetic moment along easy axes have been calculated by Kubo-Bastin formalism as implemented in KKR Green's function method[15][16]. Specifically, for the self-consistent calculations, a cutoff $l_{max}=3$ had been adopted for the angular momentum expansion, and the Perdew-Burke-Ernzerhof (PBE) type of generalized gradient approximation (GGA) is employed for describing exchange-correlation potential. The electric field induced torkances are calculated by using $10^7$ *k*-points in Brillouin Zone and 32 energy points by considering phonon scattering at room temperature.

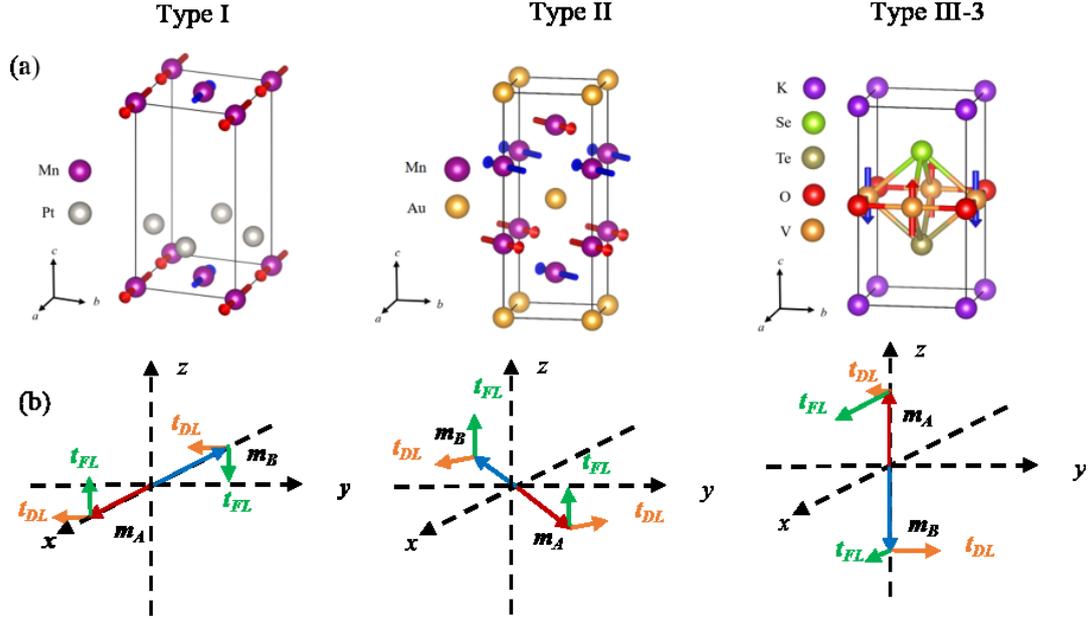

Fig. 2. (a) The crystal structure of MnPt bilayer with magnetic moments along [100] direction. The MSG of MnPt bilayer is *P4bm*. The crystal structure of $Mn_2Au$ with magnetic moments along [110] direction, and its MSG is *Fm'mm*. The crystal structure of $KV_2SeTeO$ with easy axis along [001] direction. The MSG of $KV_2SeTeO$ is *P4'm'm*. (b) The schematic of *T*-even and *T*-odd SOT on A and B spin sublattices for MnPt bilayer, $Mn_2Au$ and $KV_2SeTeO$, when electric field is applied along [100] direction.

Table II. The calculated electric field induced onsite torkance, effective field in MnPt bilayer, $Mn_2Au$ and $KV_2SeTeO$. The units for magnetic moment, torkance and effective magnetic field $B_E$ are in Bohr magneton $\mu_B$, $ea_0$ ($e$ is elementary charge, $a_0$ is Bohr radius), and $mT/(10^7 \text{ A/cm}^2)$ respectively. Based on the calculated torkance, the effective magnetic field can be evaluated as $B_E = t/(m\sigma)$, where $m$ is magnetic moment of spin sublattice and $\sigma$ is the electric conductivity of the AFM.

| Materials | MPG | Magnetic moment | Spin sublattices | $t_{even}$ | $t_{odd}$ | $B_{E,odd}$ | $B_{E,even}$ |
|---|---|---|---|---|---|---|---|
| MnPt bilayer | Type I | 4.029 | Mn(+) | 0.134 | 4.144 | 0.664 | 20.539 |
| | | | Mn(-) | 0.134 | -4.144 | -0.664 | 20.539 |
| $Mn_2Au$ | Type II | 3.827 | Mn(+) | 0.082 | 0.790 | 0.322 | 3.117 |

|  |  |  | Mn(-) | -0.082 | 0.790 | 0.322 | -3.117 |
|  |  |  | V(+) | -0.00694 | 0.0372 | -1.093 | 5.857 |
| KV$_2$SeTeO | Type III-3 | 1.077 | V(-) | 0.0114 | 0.0118 | -1.795 | -1.859 |

MnPt bilayer belongs to the type I AFM, which possesses $T\tau$ but without inversion symmetry in magnetic space group (MSG). As shown in Table II, the calculated $T$-even torques on two opposite Mn spin sublattices A and B have the same magnitude and sign, while the $T$-odd torques have the opposite sign[17]. This is similar to the spin Hall induced torque in "heavy metal/AFM" bilayers[1], where the damping like (DL, $T$-even) torque can be expressed as ***m***×(***m***×***ξ***) and the field like (FL, $T$-odd) torque has the form of (***m***×***ξ***). The spin-polarization on A and B sublattices have the same sign $\xi_A=\xi_B$.

Mn$_2$Au belonging to the aforementioned type II AFMs, possesses $PT$ but without inversion and $T\tau$ symmetry in magnetic space group (MSG). The two opposite Mn spin sublattices A and B are connected by inversion. Therefore, the calculated $T$-even SOTs have opposite sign, while the $T$-odd SOTs have same sign on Mn of A and B sublattices, which is in contrast to MnPt bilayer. In this case, the spin-polarization on A and B sublattices have the opposite sign $\xi_A=-\xi_B$. It is worth noting that the staggered $T$-even magnetic field is as large as several mT per $10^7$ A/m$^2$, which may effectively switch the Néel vector as we will discuss later.

As shown in Fig. 2, KV$_2$SeTeO belongs to the MPG-III-3 without $PT$ and $T\tau$ symmetry in its magnetic space group (MSG). The relations of $T$-even and $T$-odd torques on two opposite spin sublattices A and B cannot be described either by eq.(3) or eq.(4). Moreover, as listed in Table II, the spin-orbit torques are unequal and asymmetric on A and B spin sublattices. The current induced $T$-odd and $T$-even magnetic fields are around several mT per $10^7$ A/m$^2$, which indicates possible switching of Néel vector.

## IV. Néel vector dynamics under current induced SOTs.

To conduct Néel vector dynamics simulations driven by SOTs, one must first derive the forms of $T$-even and $T$-odd onsite torques in bulk collinear AFMs. For MnPt bilayer and Mn$_2$Au, the nonmagnetic site point groups are $4mm$. The lowest order $T$-odd torque ***T***$_{FL}$ (subsequently referred as field like torque) on each sublattice can be written in terms of magnetic moment direction **m** and in-plane charge current density $\mathbf{J}=(J_x, J_y, 0)$ as follows (see supplementary Note 1 for the details):

$$\frac{\mathbf{T}_{FL}}{M_s} = B_{FL}\mathbf{m}\times(\mathbf{z}\times\mathbf{J}) \quad (7)$$

where $M_s$ is the saturation magnetization, $B_{FL}$ is the FL torque coefficient, and **z** is the direction perpendicular to the current plane.

Similarly, the lowest order *T*-even torque $\mathbf{T}_{DL}$ (subsequently referred as damping like torque) on each sublattice can be expressed as:

$$\frac{\mathbf{T}_{DL}}{M_s} = B_{DL}\mathbf{m}\times(\mathbf{m}\times(\mathbf{z}\times\mathbf{J})) \quad (8)$$

where $B_{DL}$ is the DL torque coefficient.

Please note that, the FL and DL torques for MnPt bilayer and $Mn_2Au$ described by eq.(7) and (8) have the same forms as the torques in "Heavy metal/Ferromagnet" bilayers[1], except that one need two sets of torque coefficient ($B_{FLA}$, $B_{DLA}$), ($B_{FLB}$, $B_{DLB}$) for opposite spin sublattices A and B. The relations between the sign of the two sets of torque coefficient on A and B have been discussed in the previous section.

For $KV_2SeTeO$, the nonmagnetic site point group is *mm*2 and the field like torque on each spin sublattice can be represented as (see supplementary Note 1 for the details):

$$\frac{\mathbf{T}_{FL}}{M_s} = B_{F1}\mathbf{m}\times(\mathbf{z}\times\mathbf{J}) + B_{F2}J_y(\mathbf{m}\times\mathbf{x}) \quad (9)$$

where the first term on the right is the conventional field like term, and the second term originates from the lower $C_{2v}$ symmetry of $KV_2SeTeO$. The damping like torque on each sublattice can be represented as:

$$\frac{\mathbf{T}_{DL}}{M_s} = B_{D1}J_x\mathbf{m}\times(\mathbf{m}\times\mathbf{y}) - B_{D2}J_y\mathbf{m}\times(\mathbf{m}\times\mathbf{x}) + m_z\mathbf{m}\times(B_{D3}J_x\mathbf{x} + B_{D4}J_y\mathbf{y}) \quad (10)$$

where the first two terms are the conventional damping like terms, and the third term originating from the $C_{2v}$ symmetry of $KV_2SeTeO$. Therefore, one need two sets of torque coefficient ($B_{F1,A}$, $B_{F2,A}$; $B_{D1,A}$, $B_{D2,A}$, $B_{D3,A}$, $B_{D4,A}$) and ($B_{F1,B}$, $B_{F2,B}$; $B_{D1,B}$, $B_{D2,B}$, $B_{D3,B}$, $B_{D4,B}$) for A and B sublattices. The FL and DL torque coefficients for the three representative AFMs including MnPt bilayer, $Mn_2Au$ and $KV_2SeTeO$ can be evaluated from the calculated torkance by setting magnetic moment along various directions (see supplementary Note 1 for the details).

In order to simulate the Néel vector dynamics in bulk collinear AFMs in the presence of current induced SOTs, the following coupled Landau-Lifshitz-Gilbert (LLG) equations for magnetic moment direction $m_A$, $m_B$ on two opposite spin sublattices A and B will be solved numerically:

$$\frac{d\boldsymbol{m}_A}{dt} = -\gamma \boldsymbol{m}_A \times \boldsymbol{B}_A + \alpha \boldsymbol{m}_A \times \frac{d\boldsymbol{m}_A}{dt} + \frac{\gamma}{M_s} \boldsymbol{T}_A$$

$$\frac{d\boldsymbol{m}_B}{dt} = -\gamma \boldsymbol{m}_B \times \boldsymbol{B}_B + \alpha \boldsymbol{m}_B \times \frac{d\boldsymbol{m}_B}{dt} + \frac{\gamma}{M_s} \boldsymbol{T}_B$$

where $\gamma$ is gyromagnetic ratio, $\alpha$ is Gilbert damping. The explicit forms of SOTs for the representative AFMs $T_{A,B}/M_s$ has just been discussed. In the above LLG equations, $\boldsymbol{B}_A$ and $\boldsymbol{B}_B$ are the effective magnetic fields experienced by A and B sublattices, which can be derived from magnetic energy density $\varepsilon$: $\boldsymbol{B}_{A,B} = -\frac{\partial \varepsilon}{M_s \partial \boldsymbol{m}_{A,B}}$. The magnetic energy density $\varepsilon$ includes exchange, anisotropy and external magnetic field energy:

$$\varepsilon = \varepsilon_{exc} + \varepsilon_{ani} + \varepsilon_{ext},$$

$$\varepsilon_{exc} = \frac{J}{V} \boldsymbol{m}_A \cdot \boldsymbol{m}_B,$$

$$\varepsilon_{ani} = \frac{1}{V} \sum_{i=A,B} [K_1 (\boldsymbol{m}_i \cdot \boldsymbol{z})^2 + K_2 (\boldsymbol{m}_i \cdot \boldsymbol{z})^4 + K_3 (\boldsymbol{m}_i \cdot \boldsymbol{x})^2 (\boldsymbol{m}_i \cdot \boldsymbol{y})^2],$$

$$\varepsilon_{ext} = -M_s (\boldsymbol{m}_A + \boldsymbol{m}_B) \cdot \boldsymbol{B}_{ext}$$

where $V$ is the volume of unit cell, $J$ is the exchange energy between two opposite spin sublattices, $K_1$, $K_2$, $K_3$ are the magnetic crystalline anisotropy energy parameters expanded up to forth order for tetragonal crystal structure. Specially, the exchange fields on A and B sublattices can be described by $\boldsymbol{B}_{A,exc} = -B_{exc} \boldsymbol{m}_B$, $\boldsymbol{B}_{B,exc} = -B_{exc} \boldsymbol{m}_A$ with exchange magnetic field $B_{exc}=J_{exc}/m$. For representative AFMs, the exchange energy $J_{exc}$, exchange field $B_{exc}$, magnetic anisotropy energy parameters as well as damping constant $\alpha$ have been calculated by first-principles calculations (please see supplementary Note 2 for the computational details) and listed in Table III.

Table III. The calculated exchange energy $J_{exc}$, exchange magnetic field $B_{exc}$, magnetic crystalline anisotropy parameters $K_1$, $K_2$, $K_3$, damping constant $\alpha$ for MnPt bilayer, Mn$_2$Au and KV$_2$SeTeO.

| AFMs | $J_{exc}$(meV) | $B_{exc}$(T) | $K_1$(meV) | $K_2$(meV) | $K_3$(meV) | $\alpha$ |
|---|---|---|---|---|---|---|
| MnPtVc | 107.946 | 485.067 | 0.327 | -0.046 | 0.027 | 0.009 |
| Mn$_2$Au | 190.44 | 855.762 | 1.160 | 0.008 | -0.006 | 0.008 |
| KV$_2$SeTeO | 47.045 | 754.967 | -0.0601 | 0.0183 | 0.0389 | 0.0125 |

Since we are interesting in the Néel vector dynamics at room temperature, the anisotropy parameters has been scaled at 300 K by the Callen-Callen relation as $K_{1,2,3}(T=300\ K)/K_{1,2,3} \sim [M(T=300\ K)/M_0]^3$ [18]. The $M(T)$ curves for MnPt bilayer, Mn$_2$Au, and KV$_2$SeTeO have been simulated, which yield Néel temperature of 626 K, 1105 K, and 546 K, respectively (supplementary Note 2). The scaling factors for magnetic anisotropy at 300 K are found to to be 0.760, 0.960, and 0.819, respectively. In the following, we first discuss the Néel vector dynamics driven by SOTs for MnPt bilayer and Mn$_2$Au. The LLG equations for both materials are as follows:

$$\frac{d\boldsymbol{m}_A}{dt} = -\gamma \boldsymbol{m}_A \times \boldsymbol{B}_A + \alpha \boldsymbol{m}_A \times \frac{d\boldsymbol{m}_A}{dt} + \gamma B_{DLA} \boldsymbol{m}_A \times (\boldsymbol{m}_A \times (\boldsymbol{z} \times \boldsymbol{J})) + \gamma B_{FLA}(\boldsymbol{m}_A \times (\boldsymbol{z} \times \boldsymbol{J}))$$

$$\frac{d\boldsymbol{m}_B}{dt} = -\gamma \boldsymbol{m}_B \times \boldsymbol{B}_B + \alpha \boldsymbol{m}_B \times \frac{d\boldsymbol{m}_B}{dt} + \gamma B_{DLB} \boldsymbol{m}_B \times (\boldsymbol{m}_B \times (\boldsymbol{z} \times \boldsymbol{J})) + \gamma B_{FLB}(\boldsymbol{m}_B \times (\boldsymbol{z} \times \boldsymbol{J}))$$

For MnPt bilayer, the four torque coefficients have been determined from the calculated torkances to be $B_{DLA}$=-0.664, $B_{FLA}$=-20.539; $B_{DLB}$=-0.664, $B_{FLB}$=-20.539, and for Mn$_2$Au, $B_{DLA}$=0.455, $B_{FLA}$=4.407; $B_{DLB}$=-0.455, $B_{FLB}$=-4.407 in the unit of mT per $10^7$ A/cm$^2$. Again, please note that for MnPt bilayer, the DL torques are the same but the FL torques are opposite, while for Mn$_2$Au, the FL torques are the same but the DL torques are opposite on the two spin sublattices.

For MnPt bilayer, as a demonstration the spin dynamics are simulated by setting electric field and initial Néel vector along [100] direction (i.e. $x$ axis). In the following, the Néel vector is defined as $\boldsymbol{l}=(\boldsymbol{m}_A-\boldsymbol{m}_B)/2$. As shown in Fig. 3(a), by applying sufficiently large charge current density for instance +$J_x$=1.1×10$^{10}$ A/cm$^2$, the Néel vector starts to oscillate in the *xz* plane but cannot be switched to another easy axis. The reason for the oscillation of Néel vector is the much larger FL torque in comparison with DL torque in MnPt bilayer. The FL torque provides a large effective magnetic field (~22.6 T) along *y* direction and enable the oscillation of Néel vector in the *xz* plane. The spin dynamics process of MnPt bilayer is dependent on the relative magnitudes of DL and FL torques. If the FL torques are reduced, the Néel vector may be switched by 90° to another easy axis. As shown in Fig.3 (b) by reducing FL torque coefficients to 10% of its original value and applying the current density $J_x=8.6\times10^9$ A/cm$^2$, the Néel vector now can be switched from *x* to *y* axis. It is worth noting that such switching is independent of the charge current direction and actually non-deterministic by considering the thermal fluctuation of initial Néel vector as we discuss below.

We then discuss the possible electric regulations of antiferromagnetic domains in MnPt bilayer with reduced FL torques. As shown in Fig.3 (d) and (e), the domains with Néel vector along *x* direction can be switched to *y* axis by applying the charge current along *x* direction. But the resultant Néel vector with the same proportions of [010] and [0-10] directions if the thermal fluctuation of initial Néel vectors are considered. Similarly, the domains with Néel vector pointing along *y* direction can be switched to *x* axis by the charge current along *y* direction, but with the same ratios of final [100] and [-100] Néel vector directions. These results indicate that in type I AFMs with the valid magnitudes of DL and FL torques, the SOTs could drive 90° switch of AFM domains by sufficient charge current along *x* and *y* axis, but the state of multiple domains cannot be regulated into single domain, and at least two domains with opposite Néel vector are remaining for the initial states with all the four possible domains.

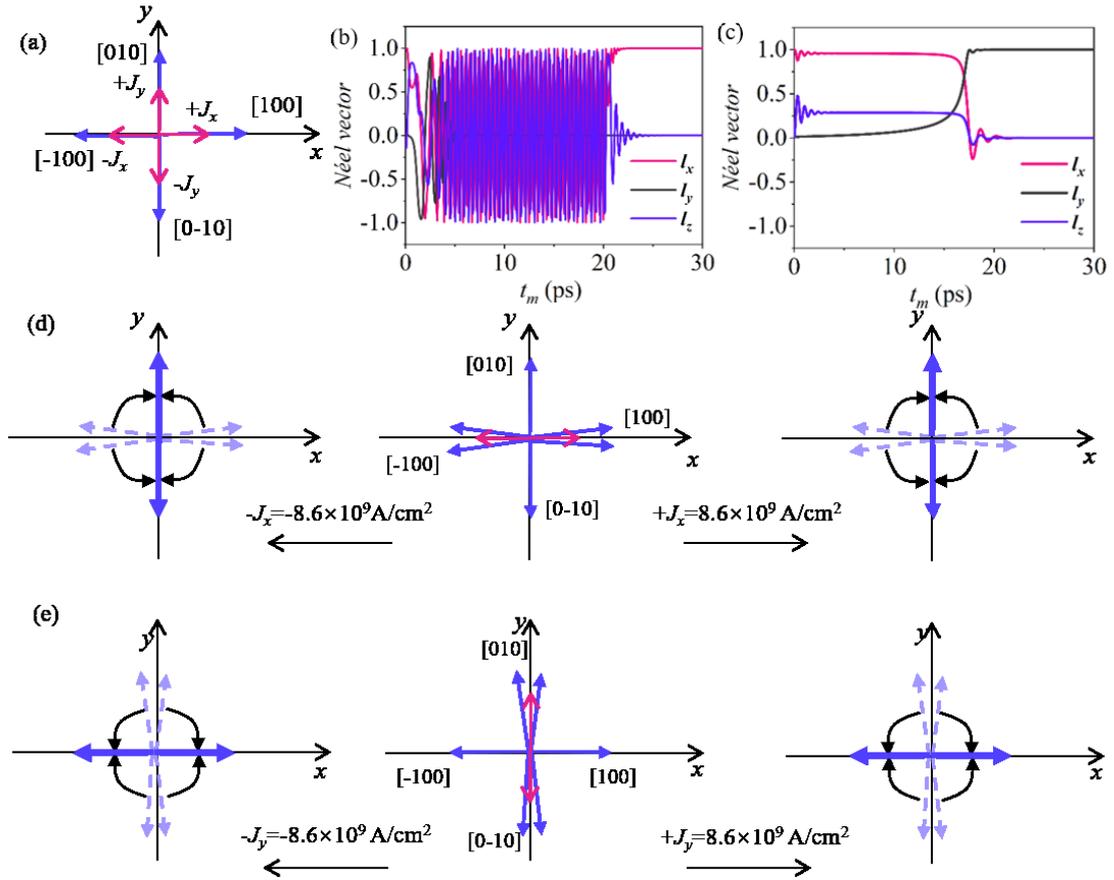

Fig. 3. (a) The four energy degenerate Néel vector orientations (blue arrows) for MnPt bilayer and the illustrations of charge current along *x* and *y* axis as indicated by red arrows. (b) The oscillation of Néel vector in MnPt bilayer by setting initial Néel vector and current density $+J_x=1.1\times10^{10}$ A/cm$^2$ along [100] direction. (c) The switching of Néel vector in MnPt bilayer from [100] to [010] by applying charge current density $+J_x=8.6\times10^{9}$ A/cm$^2$ with reducing FL torques to 10% of its original value. The simulation time is 80 ps and the current pulse time is 20 ps. (d) The schematics of electric modulations of AFM domains by applying charge current along *x* direction as indicated by red arrows. In middle panel, the initial Néel vectors along [100] and [-100] directions are shown by the blue arrows indicating the possible configurations by considering thermal fluctuation. (e) The same as (d), except that the charge current is applied along *y* direction. In (d) and (e) the FL torques are reduced to 10% of its original value.

The above discussed spin dynamics of MnPt bilayer are driven by the DL torques instead of FL torques. By using the similar forms of torque as in MnPt bilayer, we are able to investigate the Néel vector dynamics in other different cases. For instance, if the magnetic easy axis of MnPt bilayer is set to be along *z* axis by changing magnetic anisotropy parameter $K_1$ to $-K_1$, the Néel vectors can oscillate but cannot be switched by current induced SOTs. In addition, the Néel vector dynamics for Type II-2 collinear AFMs may also be investigated by setting opposite DL torques on A and B sublattices in MnPt bilayer. It is found that in this case, neither Néel vector oscillation nor switching is possible, which demonstrates that in Type III-2 collinear AFMs without net DL and FL torques, the desirable antiferromagnetic spin dynamics driven by SOT are not favorable.

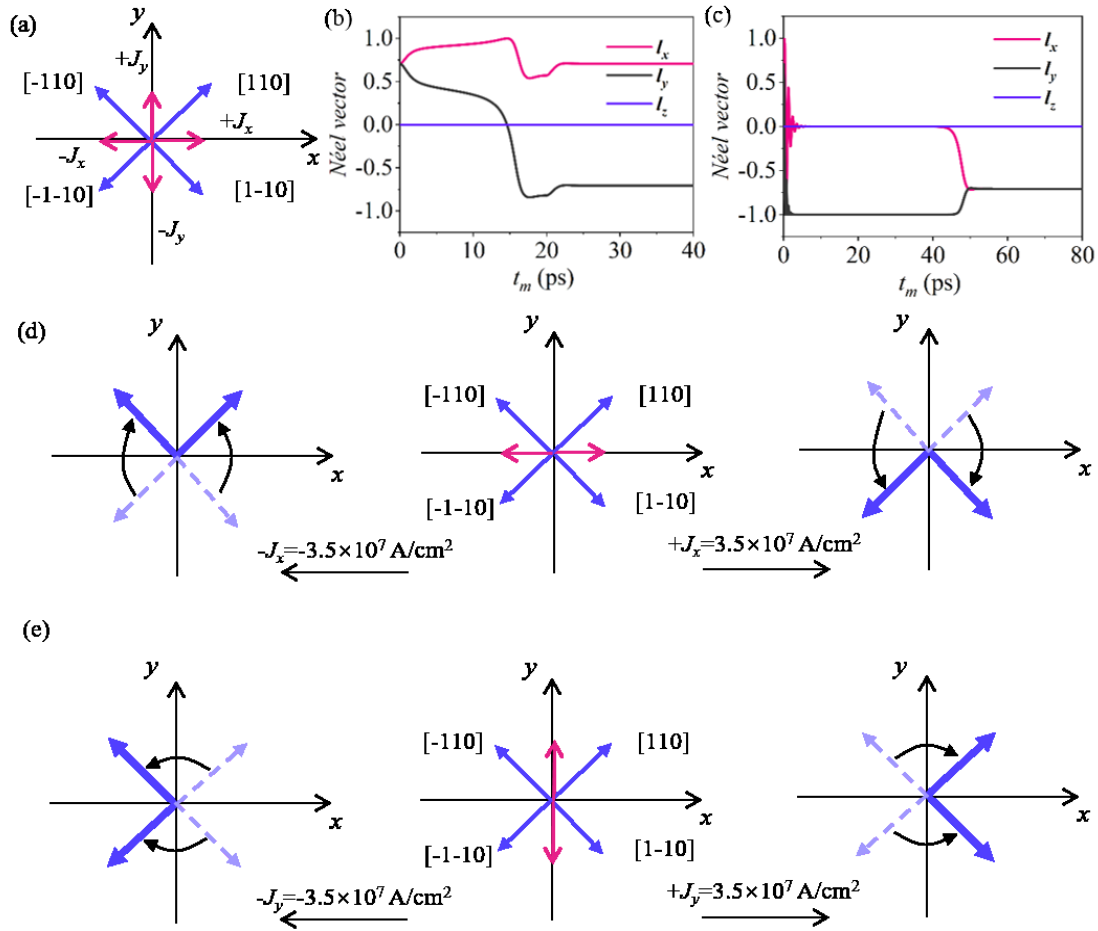

Fig. 4. (a) The four energy degenerate Néel vector orientations (blue arrows) for Mn$_2$Au, and the illustrations of charge current along $x$ and $y$ axis (red arrows). (b) The 90° switching of Néel vector from initial [110] to [1-10] direction by applying $+J_x=3.5\times 10^7$ A/cm$^2$. (c) The 180° switching of Néel vector from initial [110] to [-1-10] by applying $+J_x=1.5\times 10^8$ A/cm$^2$. The simulation time is 80 ps and the current pulse time is 20 ps. (d) The schematics of electric modulations of AFM domains by applying charge current along $x$ direction. In the middle panel, the possible initial Néel vectors in different AFM domains are indicated by blue arrows and the current directions are shown by the red arrows. (e) The same as (d), except that the charge current is applied along *y* direction.

For $Mn_2Au$ as shown in Fig. 4(a), there are four energy degenerate Néel vector directions and the deterministic switching can be achieved by SOTs ("deterministic" means the initial Néel vector can be switched to a definite final state by considering stochastic thermal magnetic fields and fluctuations of Néel vector). As depicted in Fig.4 (b), the initial Néel vector along [110] can be switched through a 90° clockwise rotation to [1-10] by applying 20 ps pulsed current with density $+J_x=3.5\times 10^7$ A/cm$^2$. It is noteworthy that the 180° switching of Néel vector may also be possible. As shown in Fig.4(c), the Néel vector can be switched from [110] to [-1-10] by applying a relatively larger charge current density $+J_x=1.5\times 10^8$ A/cm$^2$. In contrast to the case of MnPt bilayer, the switching process of $Mn_2Au$ have been driven by the FL torques, instead of DL torques. The FL torques on two opposite Mn sublattices in $Mn_2Au$ have the same sign and magnitude, which enable efficient Néel vector switching with relatively low current density.

There are four energy degenerate Néel vector orientations [110], [1-10], [-1-10], [-110], *i.e.* four possible antiferromagnetic domains in $Mn_2Au$. By further considering the applied charge current directions $\pm J_x$, $\pm J_y$, there are in total 16 configurations for investigating SOT induced switching. However, due to the presence of rotation symmetry of $Mn_2Au$, there are only two independent configurations $\{[110], +J_x\}$ and $\{[110], -J_x\}$ that are needed to be systematically investigated. The SOT induced Néel vector dynamics in bulk AFMs are dependent both on the current density and the length of pulsed current. The current density and the length of current pulse dependence of switching for those two configurations has been listed in details in supplementary Table S1 and S2.

As shown in Fig.4 (d), if all the four energy degenerate AFM domains exist in $Mn_2Au$, by applying charge current along *x* direction $+J_x=3.5\times 10^7$ A/cm$^2$, the [110] and [-110] domains will be electrically switched to [1-10] and [-1-10] respectively, whereas [1-10] and [-1-10] domains will not change. There will be two remaining AFM domains with

Néel vectors pointing along [1-10] and [-1-10]. In contrast to MnPt bilayer, the switching paths of Néel vector in Mn$_2$Au depend on the direction of current. By applying -$J_x$=-3.5×10$^7$ A/cm$^2$, the remaining two AFM domains are [110] and [-110]. For the current in the $y$ direction, as shown in Fig.4 (e), the results are analogous. Importantly, by applying current in the $x$ and $y$ directions sequentially, the multiple domains with various Néel vector orientations can be modulated into single domain with Néel vector along preset direction. The combinations of (+$J_x$, +$J_y$), (+$J_x$, -$J_y$), (-$J_x$, +$J_y$), (-$J_x$, -$J_y$) will set the multiple domain states into single domain with Néel vectors along [1-10], [-1-10], [110], [-110] directions respectively. What's more, by applying currents $J_x$ and $J_y$ sequentially, the Néel vector can be switched by 180°. For instance, the single domain with Néel vector along [110] direction can be switched to [-1-10] by applying +$J_x$ and -$J_y$ sequentially. In all, for Mn$_2$Au, by applying $J_x$ or $J_y$ the deterministic 90° switch can be easily achieved, the multiple domains state may be written into two states, while by applying two current $J_x$ and $J_y$ sequentially, the multiple domains state may be written into single state and the 180° switch can be achieved.

The easy axis of KV$_2$SeTeO lies along $z$ direction, and the non-conventional DL torques are involved. The LLG equations for KV$_2$SeTeO when charge current is along $x$ axis is as follows (please see supplementary Note 1 for the detailed derivations of SOTs in KV$_2$SeTeO):

$$\boldsymbol{T}_{FL}/M_s = B_{F1}J_x(\boldsymbol{m}\times\boldsymbol{y}); \quad \boldsymbol{T}_{DL}/M_s = B_{D1}J_x\boldsymbol{m}\times(\boldsymbol{m}\times\boldsymbol{y}) + B_{D3}J_x m_z(\boldsymbol{m}\times\boldsymbol{x})$$

$$\frac{d\boldsymbol{m}_A}{dt} = -\gamma\boldsymbol{m}_A\times\boldsymbol{B}_A + \alpha\boldsymbol{m}_A\times\frac{d\boldsymbol{m}_A}{dt} + \gamma\{B_{D1,A}J_x\boldsymbol{m}_A\times(\boldsymbol{m}_A\times\boldsymbol{y}) + B_{D3,A}J_x(\boldsymbol{m}_A\cdot\boldsymbol{z})(\boldsymbol{m}_A\times\boldsymbol{x})\}$$
$$+\gamma B_{F1,A}J_x(\boldsymbol{m}_A\times\boldsymbol{y})$$

$$\frac{d\boldsymbol{m}_B}{dt} = -\gamma\boldsymbol{m}_B\times\boldsymbol{B}_B + \alpha\boldsymbol{m}_B\times\frac{d\boldsymbol{m}_B}{dt} + \gamma\{B_{D1,B}J_x\boldsymbol{m}_B\times(\boldsymbol{m}_B\times\boldsymbol{y}) + B_{D3,B}J_x(\boldsymbol{m}_B\cdot\boldsymbol{z})(\boldsymbol{m}_B\times\boldsymbol{x})\}$$
$$+\gamma B_{F1,B}J_x(\boldsymbol{m}_B\times\boldsymbol{y})$$

The DL and FL torque coefficients for A and B sublattices have been determined from the calculated torkances to be $B_{D1,A}$=3.251, $B_{D3,A}$=1.897, $B_{F1,A}$=-5.857 and $B_{D1,B}$=-2.568, $B_{D3,B}$=-1.922, $B_{F1,B}$=1.859 in the unit of mT per 10$^7$ A/cm$^2$.

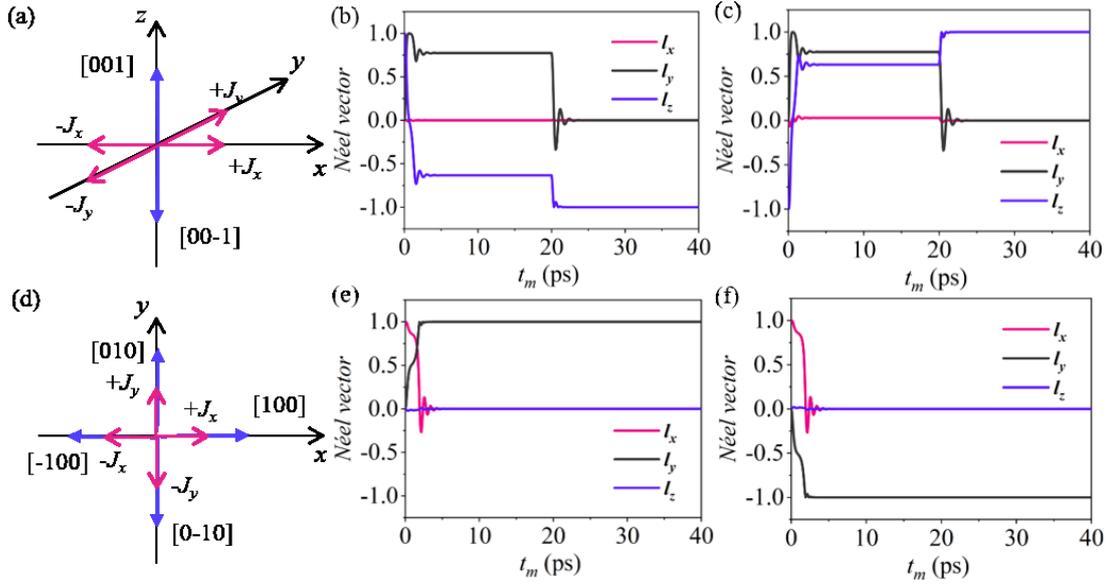

Fig. 5. (a) The two energy degenerate Néel vector orientations (blue arrows) for $KV_2SeTeO$ with easy axis along $z$ direction and the illustrations of charge current along $x$ and $y$ axis (red arrows). (b) The 180° switching of Néel vector by applying current density $+J_x=2.4\times10^9$ A/cm$^2$ from [001] to [00-1]. (c) The 180° switching of Néel vector by applying current density $+J_x=2.4\times10^9$ A/cm$^2$ from [00-1] to [001]. (d) The four energy degenerate Néel vector orientations for $KV_2SeTeO$ with in-plane anisotropy by setting anisotropy parameter $K_1$ to $-K_1$. (e) The 90° switching of Néel vector from initial [100] to [010] direction by applying $+J_x=6.7\times10^8$ A/cm$^2$. (f) The 90° switching of Néel vector from initial [100] to [0-10] direction by applying $-J_x=-6.7\times10^8$ A/cm$^2$. The simulation times are 80 ps and current pulse time are 20 ps for all the cases.

The 180° deterministic switching in inversion symmetry breaking altermagnet $KV_2SeTeO$ with perpendicular magnetic anisotropy (PMA) can be achieve by SOTs. As shown in Fig.5 (a) and (b), the Néel vector can be switched from [001] to [00-1] by applying charge current $J_x=2.4\times10^9$ A/cm$^2$, and it can flip back by applying the same current density. Similar to $Mn_2Au$, instead of DL torques, such switching is driven by FL torques with the same sign (although different magnitude) on the two opposite V spin sublattices. The simulated spin dynamic process for $KV_2SeTeO$ with initial Néel vector along [001] as a function of current density along $x$ axis $J_x$ and the length of pulsed current has been listed in the Table S3 in supplementary materials. It is shown that there exists a current density window (232.4~356.1×10$^7$ A/cm$^2$) for the 180° deterministic switching. In addition, it is found that the switching window becomes narrow when the PMA of $KV_2SeTeO$ decrease. Our results indicate that the switching window disappears when magnetic anisotropy parameters $K_1$, $K_2$, $K_3$ are reduced to 10% of the original values.

By setting the magnetic anisotropy parameter $K_1$ to be $-K_1$, the easy axis of $KV_2SeTeO$ now can be turned to $x$, $y$ axes, and we could examine the switching process of type III-3 AFMs with easy plane magnetic anisotropy. As shown in Fig. 5 (e), by applying charge current with density $+J_x=6.7\times10^8$ A/cm$^2$, the initial Néel vector along [100] can be deterministically switched by 90° to [010] and it can be switched back to [100] by applying current density $+J_y=4.2\times10^8$ A/cm$^2$. The simulated spin dynamics process for $KV_2SeTeO$ with initial Néel vector along [100] and [010] as a function of current density $J_x$ and $J_y$ respectively and the length of pulsed current has been listed in the Table S5 and S6 in supplementary materials.

Electric regulation of domains in type III-3 AFM $KV_2SeTeO$ is also manifest. To be consistent with previous MnPt bilayer and $Mn_2Au$, we discuss the case of $KV_2SeTeO$ with magnetic easy axes in $x$ and $y$ directions. Suppose we have AFM domains with the same proportion along [100], [010], [-100] and [0-10] directions, by applying charge current $+J_x=6.7\times10^8$ A/cm$^2$, the Néel vectors in three-quarters of the domains will point along [010] direction and the remaining one-quarter of the domain along [0-10], which already regulate the Néel vectors to the prefer direction. Similar to $Mn_2Au$, the electric writing of multiple AFM domains into single domain state is feasible by applying $J_x$ and $J_y$ current sequentially. The combinations of $(+J_x, +J_y)$, $(+J_x, -J_y)$, $(-J_y, +J_x)$, $(+J_y, -J_x)$ will set the multiple domain states into single domain with Néel vector pointing along [100], [-100], [010], [0-10] directions respectively. What's more, by applying $J_x$ and $J_y$ sequentially, the Néel vector can be switched by 180°. For instance, the single domain with Néel vector along [100] can be switched to [-100] direction by applying $+J_x$ and $-J_y$ sequentially.

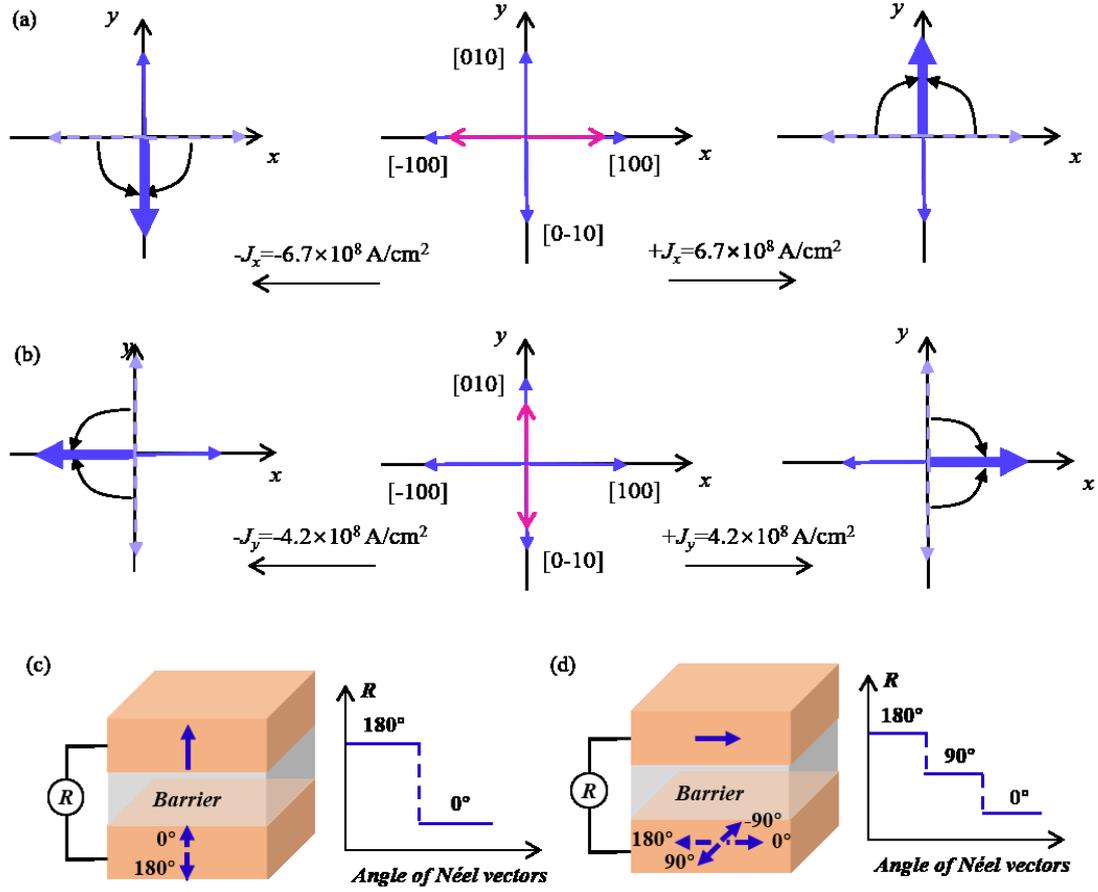

Fig. 6 (a) The schematics of electric modulations of AFM domains by applying charge current along *x* direction. In the middle panel, the possible initial Néel vectors in different AFM domains are indicated by blue arrows and the current directions are shown by the red arrows. (b) The schematics of electric modulations of AFM domains by applying charge current along *y* direction. (c) and (d) The schematic diagrams for SOT switching of Néel vectors in type III-3 AFMs and the electric reading by TMR in antiferromagnetic tunnel junctions.

The deterministic switching of Néel vectors and electric regulation of the domains in type III-3 AFMs by SOTs are important. It is known that type III AFMs exhibit spin-splitting in momentum space and *T*-odd spin transport, which enable electric detection of Néel vectors via magnetoresistance effect[19][20][21]. The spin-polarized band structure of $KV_2SeTeO$ has been shown in Fig. S2 in supplementary Note 4. As depicted in Fig.6 (c) and (d), the reorientation of Néel vectors in type III-3 AFMs with PMA and easy plane anisotropy can be detected by tunneling magnetoresistance effect (TMR) in antiferromagnetic tunnel junctions by using type III-3 AFMs as antiferromagnetic free layer. The fully electric writing and reading of antiferromagnetic states is precisely the ultimate objective we aim to realize for antiferromagnetic memory applications.

## V. Summary

In summary, we classify the SOTs in collinear AFMs into different types and among them three types exhibiting nonzero onsite damping and field like torques on two spin sublattices as well as net torques represented by MnPt bilayer (type I), $Mn_2Au$ (type II), and $KV_2SeTeO$ (type III-3). The onsite torques for those AFMs have been calculated by first-principles calculations and the forms of torques by considering different crystal symmetries have been derived. The SOTs induced dynamics are simulated based on LLG equations. For MnPt bilayer with net DL torque, SOTs can only trigger the Néel vector oscillation and non-deterministic switching. In contrast, the deterministic Néel vector switching can be achieved in $Mn_2Au$ and $KV_2SeTeO$ driven by FL torques. By using two writing current paths, in type II and type III-3 AFMs, the SOTs enable 180° switching and electric writing of single AFM domain. Finally, we need to mention that, in this work the switching of Néel vectors are realized by rotation, the domain wall motion is not involved in the switching process. In practical, the current density may be one order of magnitude smaller when the domain wall motion mechanism plays an important role in the switching process[22][23] which will be deferred to our future research. Our work should deepen our comprehension of SOTs and the Néel vector dynamics in AFMs, and may inspire experimental investigations and promote related device applications.

## Acknowledgement

This work was supported by the National Natural Science Foundation of China (Grants No. T2394475, No. T2394470, No. 12174129).